\documentclass[12pt]{article}
\usepackage[english]{babel}
\usepackage{epsfig}
\begin{document}

\noindent {\bf About $K^o, \bar K^o$ meson oscillations at
strangeness violation by weak interactions without and with taking
into account meson decays}

\par
\begin{center}
Kh. M. Beshtoev
\par
\vspace{0.3cm} Joint Institute for Nuclear Research, Joliot Curie
6, 141980 Dubna, Moscow region, Russia.
\end{center}

Abstract

\par
This work considers $K^0, \bar K^0$ meson mixings and oscillations
via $K^0_1, K^0_2$ meson states at strangeness violation by weak
interactions in two cases - without and with taking into account
$K^0_1, K^0_2$ meson decays. In the first case the oscillation
theory correctly describes $K^o, \bar K^o \to K^o, \bar K^o$
transition processes. In the second case, when we take into
account meson decays, the oscillation theory cannot correctly
describe this process at long distances since mainly $K^o_2$
mesons remain at long distances therefore the condition for
oscillations is absent. So, at long distances in this process only
long living $K^0_2$ mesons are present but not $K^o, \bar K^o$
mesons.\\


\section{Introduction}

\par
Oscillations of $K^o$ mesons (i.e., $K^{o} \leftrightarrow \bar
K^{o}$) were theoretically \cite{1} and experimentally \cite{2}
investigated in the 50-s and 60-s. Recently an understanding has
been achieved that these processes go as a double-stadium process
\cite{3,4,5,6}. A detailed study of $K^{o}$ meson mixing and
oscillations is very important since the theory of neutrino
oscillations is built in analogy with the theory of $K^o$ meson
oscillations.
\par
It is necessary to remark that in literature devoted to this
subject $K^o_1, K^o_2$ mesons are seldom mentioned. These mesons
appear at violation of strangeness -$S$ (see \cite{7}). However,
taking into account these states is very important since the weak
interaction process with $S$ violation is faster than the weak
interaction process with $CP$ violation, i.e., first the $K^o_1,
K^o_2$ mesons are produced and then - the $K_S, K_L$ meson states.
It is well seen from a very small probability of $CP$ violation in
the system of $K^o$ mesons. We cannot correctly understand the
$K^o$ processes if we do not take into account the presence of
$K^o_1, K^o_2$ meson states.
\par
A phenomenological analysis of $K^{o}$ meson processes was done in
\cite{8, 9}. In this work another approach is used to consider
$K^{o}$ meson processes. This work is based on the principles of
the quantum field theory or particle physics. It is supposed that
particles ($K^o$ mesons) while production have no widths for
decomposition, i.e., they can only decay in a usual way, as is in
the case of particle physics. This remark is important since in
this case particles cannot form wave packets and the wave packets
then can be formed only from a big number of identical particles
(mesons). The assumption that $K^o$ mesons can be considered as
wave packets is a hypothesis and has at present neither
experimental nor theoretical confirmation. But at the same time,
from our experience in particle physics we can make a conclusion
that elementary particles have no widths to consider them as wave
packets.
\par
At first we consider mixings and oscillations of $K^{o}, \bar
K^{o}$ mesons at violation of strange\-ness - $S$ in the weak
interactions without taking into account the decay widths of
$K^o_1, K^o_2$ mesons and then we consider mixings and
oscillations of $K^{o}, \bar K^{o}$ mesons taking into account the
decay widths of $K^o_1, K^o_2$ mesons.

\section{Vacuum mixings and oscillations of $K^o, \bar K^o$
mesons at strangeness violation by weak interactions without
taking into account decay widths}

At strangeness violation the $K^o, \bar K^o$ states are
transformed into superposition states of $K^o_1, K^o_2$ meson
states:
$$
\begin{array} {cc} K^o = {cos \theta K^o_1 + sin \theta K^o_2} , \\ \bar K^o
= {- sin \theta K^o_1 + cos \theta \bar K^o_2} .\end{array}
\eqno(3)
$$
and the inverse transformation gives:
$$
\begin{array} {cc} K^o_1 = {cos \theta K^o - sin \theta \bar K^o} , \\ K^o_2
= { sin \theta K^o + cos \theta \bar K^o}. \end{array} \eqno(4)
$$
For $CPT$ invariance of the weak interactions $m_{K^{o} K^{o}} =
m_{\bar K^o \bar K^o}$ then  mixing angle $\theta$ will be equal
to ${\pi \over 4}$. Then from expressions (3) and (4) we get the
following:
$$
K^o = \frac{K^o_1 +  K^o_2}{\sqrt{2}}, \qquad \bar K^o =
\frac{K^o_1 - K^o_2}{\sqrt{2}} ; \eqno(5)
$$
$$
K^o_1 = {{K^o - \bar K^o}\over {\sqrt{2}}} , \qquad K^o_2 = {{K^o
+ \bar K^o}\over {\sqrt{2}}} . \eqno(5')
$$

\par
The length of $K^{o}$ meson oscillations at low velocities $v$ is:
$$
L = vt = {{2\pi v} \over {\mid m_1 - m_2 \mid}} = \frac {{2\pi
p_{K^o}}} {2 m_{K^o} 2 m_{K^o \bar K^o}} = \frac {{2\pi p_{K^o}}}
{(m_2^2 - m_1^2)}. \eqno(6)
$$
In the standard approach \cite{8}, \cite{9} \cite{10}to
$K^{o}$meson oscillations it is supposed that $K^{o}$ mesons are
produced at once in the form of superposition states (4). It means
that while producing $K^{o}$, $\bar K^o$ mesons their mass matrix
has a nondiagonal form. In order to find their eigenstates, we
have to diagonalize this matrix. Then we see that their
eigenstates are $K^{o}_1$-, $K^o_2$ mesons, i.e., this case has to
produce $K^{o}_1$-, $K^o_2$ mesons but not $K^{o}$, $\bar K^o$
mesons.
\par
As a matter of fact since $K^{o}$ mesons are eigenstates of strong
interactions they cannot be produced in superposition states of
$K^{o}_{1}$-, $K^{o}_{2}$ mesons. $K^{o}$ mesons become
superposition states of $K^{o}_{1}$-, $K^{o}_{2}$ mesons when weak
interactions transform them in a superposition of eigenstates. It
is important to note that the weak interactions in contrast to the
strong interactions will produce $K^{o}_{1}$, $K^{o}_{2}$ meson
states. Now we come to a detailed consideration of $K^{o}$ meson
oscillations in the framework of the mass mixing scheme.
\par
The mass matrix of $K^{o}$ mesons has the following form:
$$
\left(\begin{array}{cc} m_{K^o}& 0 \\ 0 & m_{\bar K^o}
\end{array} \right) . \eqno(7)
$$
Strangeness is violated due to the weak interactions and in this
mass matrix nondiagonal terms appear, then it gets the following
form ($CP$ is conserved):
$$
\left(\begin{array}{cc}m_{K^o} & m_{K^o \bar K^o} \\ m_{\bar K^o
K^o} & m_{\bar K^o} \end{array} \right) . \eqno(8)
$$
At diagonalization of this matrix we obtain $K^{o}_{1}, K^{o}_{2}$
meson states and the states $K^o, \bar K^o$ are transformed in
superposition states of $K^{o}_{1}, K^{o}_{2}$ states (see
expression (4)). Expressions for their angle mixing and masses are
given with expressions (10)-(12) in work \cite{11}.
\par
The expression for $sin^2 2\theta$ is given by ($\theta$ is the
angle of mixing):
$$
sin^2 2\theta = \frac{(2m_{K^o \bar K^o})^2} {(m_{K^o} - m_{\bar
K^o})^2 +(2m_{K^o \bar K^o})^2} , \quad \left(\begin{array}{cc} m_{K^o_1} & 0 \\
0 & m_{K^o_2} \end{array} \right) .\eqno(9)
$$
The evolution of $K^o _{1}, K^o _{2}$ meson states with masses
$m_{1}, m_{2}$ will be given with the following expression:
$$
K^o _{1}(t) = e^{-i E_1 t} K^o _{1}(0), \qquad K^o _{2}(t) = e^{-i
E_2 t} K^o _{2}(0) , \eqno(10)
$$
where
$$
E^2_{k} = (p^{2} + m^2_{k}), k = 1, 2  .
$$
\par
If these mesons are moving without interactions, then
$$
\begin{array}{c}
K^o(t) = cos \theta e^{-i E_1 t} K^o_{1}(0) + sin \theta
e^{-i E_2 t} K^o_{2}(0) , \\
\bar K^o(t) = - sin \theta e^{-i E_1 t} K^o_{1}(0) + cos \theta
e^{-i E_2 t} K^o_{2}(0) .
\end{array}
\eqno(11)
$$
Using expression (3) for $K^o _{1}$ и $K^o _{2}$ and putting them
in (11), we obtain
$$
K^o (t) = \left[e^{-i E_1 t} cos^{2} \theta + e^{-i E_2 t} sin^{2}
\theta \right] K^o(0) +
$$
$$
+ \left[e^{-i E_1 t} - e^{-i E_2 t} \right] sin \theta \cos \theta
\bar K^o(0) , \eqno(12)
$$
$$
\bar K^o(t) = \left[e^{-i E_1 t} sin^{2} \theta + e^{-i E_2 t}
cos^{2} \theta \right] \bar K^o(0)  +
$$
$$
+ \left[e^{-i E_1 t} - e^{-i E_2 t} \right] sin\theta cos \theta
\bar K^o(0) .
$$
\par
The probability that meson $K^o$ produced in moment $t = 0$ will
be in moment $t \neq 0$ in state $\bar K^o(0)$ meson, is given by
a squared absolute value of the amplitude in (12), i.e.,
$$
\begin{array}{c}
P(K^o \rightarrow \bar K^o) = \mid(\bar K^o(0) \cdot K^o(t))\mid^2 =\\
 = {1\over 2} \sin^{2} 2\theta \left[1 - cos ((E_2 - E_1) t)
\right] \equiv \frac{1}{2} \left[1 - cos ((E_2 - E_1) t) \right] ,
\end{array}  \eqno(13)
$$
where $\theta = \pi/4$. And the probability that meson $K^o$
produced at moment $t = 0$ will be at moment $t \neq 0$ in the
state of $K^o$ meson, is given by expression
$$
\begin{array}{c}
P(K^o \rightarrow K^o) = \mid(\bar K^o(0) \cdot K^o(t))\mid^2 =\\
 = {1\over 2} \sin^{2} 2\theta \left[1 + cos ((E_2 - E_1) t)
\right] \equiv \frac{1}{2} \left[1 + cos ((E_2 - E_1) t) \right].
\end{array}  \eqno(14)
$$
From expressions (13) and (14) we see that at $(E_2-E_1) t = 2 \pi
n$ there will be only $K^o$ state and at $(E_2-E_1) t = \pi n$
there will be only $\bar K^o$ state (where $n$ is an integer
number). If there is no $K^o$ meson decay then these oscillations
will continue endlessly. Below we will consider the case when
mesons will decay.
\par
Then in the ultrarelativistic case the length $L_{1 2}$ of $K^o,
\bar K^o$ meson oscillations is as follows:
$$
L_{1 2} = \frac{\gamma}{2 \Delta} \equiv \frac{2 \pi h c \gamma}{2
\Delta} . \eqno(15)
$$
And the numerical value for the length of $R_{2 1}$  of $K^0, \bar
K^0$ oscillations is
$$
R_{2 1} \cong \frac{\gamma}{2 \Delta} \equiv \frac{2 \pi h c
\gamma}{2 \Delta} = 0.352 \gamma [m], \eqno(16)
$$
where $\gamma$ is a usual relativistic factor.
\par
Figure 1 gives probabilities of primary $K^o$ meson transitions in
vacuum without taking $K^o_1, K^o_2$ decays into account.
\par
It is necessary to remark that $CP$$K^o_1 = K^o_1$ and $CP K^o_2 =
-K^o_2$, i.e., $CP$ parity $K^o_1$ meson is a positive value and
it can decay into two $\pi$ mesons, and $CP$ parity of $K^o_2$
meson is a negative value and it can decay into three $\pi$
mesons. The ratio between the life time of $K^o_2$ and $K^o_1$ is
$\frac{\tau_{K_2}}{\tau_{K_1}} = 580$. The case of $CP$ violation
will be considered in another work.\\

\newpage
\begin{figure}[h!]
\begin{center}
\includegraphics[width=8.0cm]{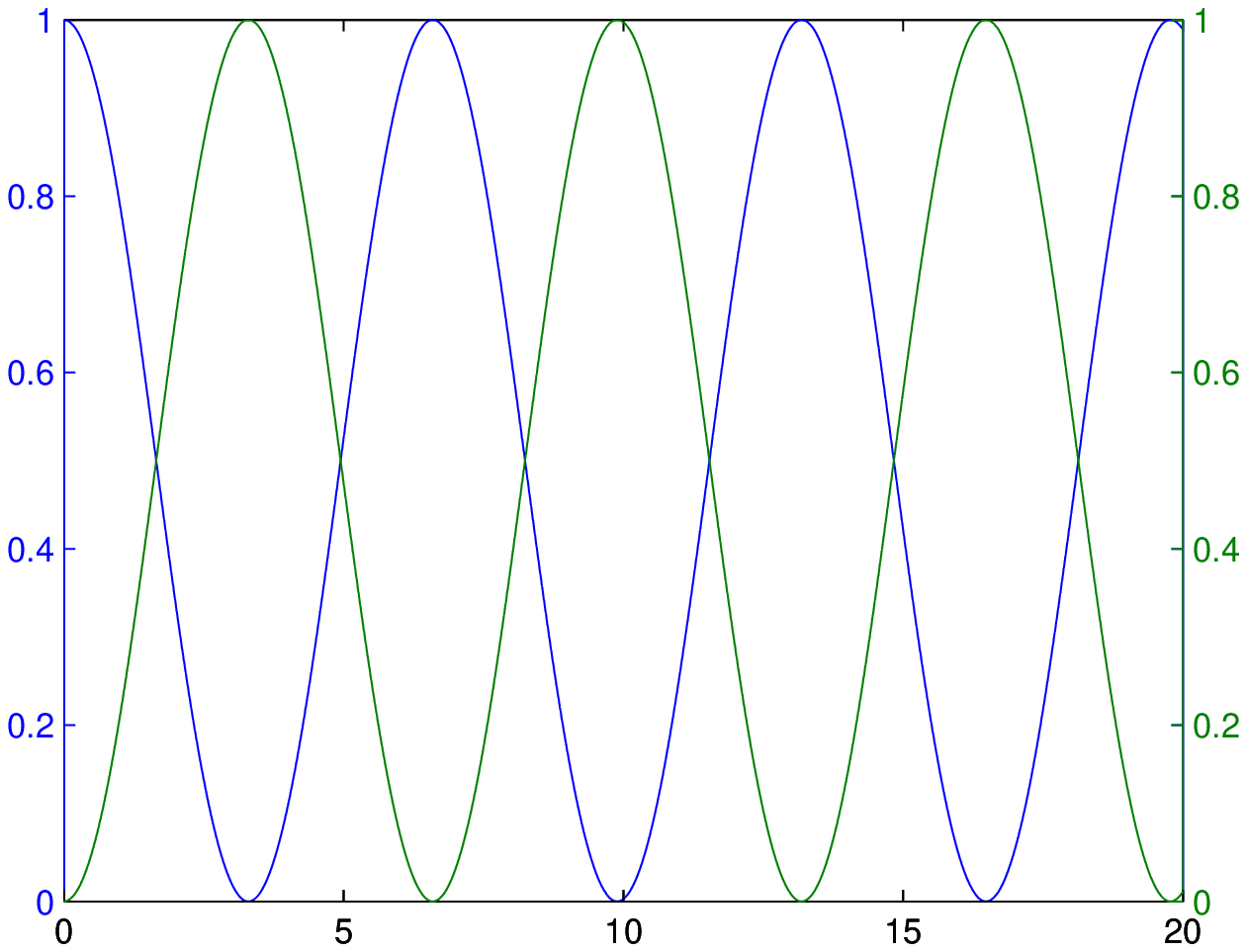}
   \end{center}
   \end{figure}

Figure 1. Probabilities of primary $K^o$ meson transitions into
$K^o, \bar K^o$ in vacuum without taking $K^o_1, K^o_2$ decays
into account ($t = \frac{t(sec.)}{\tau_{k^0_2}} \cong \frac{t(sec.)}{\tau_{k_S}}$).\\

\section{Vacuum mixings and oscillations of $K^o, \bar K^o$
mesons at strangeness violation by weak interactions taking into
account decay widths}

\par
Taking into account that $K^o_1, K^o_2$ decay and have decay
widths - $\Gamma_1, \Gamma_2$, we get that $K^o _{1}, K^o _{2}$
mesons with masses $m_{1}$, $m_{2}$ will evolve in dependence on
time according to the following law \cite{12}:
$$
K^o _{1}(t) = e^{-i E_1 t - \frac{\Gamma_1 t}{2}} K^o _{1}(0),
\qquad K^o _{2}(t) = e^{-i E_2 t - \frac{\Gamma_2 t}{2}} K^o
_{2}(0) , \eqno(17)
$$
$E_2 - E_1$ is given by expression (25) in \cite{11} and it is
equal to $\frac{2 m_{K^o} \Delta}{E_{K^o}}$
$$
E_2 - E_1 \simeq \frac{2 m_{K^o} \Delta}{E_{K^o}} . \eqno(18)
$$
In this work we suppose that $\Gamma_k = \gamma \Gamma^o_k$, where
$\Gamma^o_k$ is $K^o_k $ meson width in rest and $\gamma =
E_k/m_k$ is a usual  relativistic factor ($k = 1,2$).
\par
Then for $K^o $ и $K^o$ we obtain
$$
K^o (t) = \left[e^{-i E_1 t - \frac{\Gamma_1 t}{2}} cos^{2} \theta
+ e^{-i E_2 t - \frac{\Gamma_2 t}{2}} sin^{2} \theta \right]
K^o(0) +
$$
$$
+ \left[e^{-i E_1 t - \Gamma_1/2} - e^{-i E_2 t - \frac{\Gamma_2
t}{2}} \right] sin \theta \cos \theta \bar K^o(0) , \eqno(19)
$$
$$
\bar K^o(t) = \left[e^{-i E_1 t - \frac{\Gamma_1 t}{2}} sin^{2}
\theta + e^{-i E_2 t - \frac{\Gamma_2 t}{2}} cos^{2} \theta
\right] \bar K^o(0)  +
$$
$$
+ \left[e^{-i E_1 t - \frac{\Gamma_1 t}{2}} - e^{-i E_2 t -
\frac{\Gamma_2 t}{2}} \right] sin\theta cos \theta \bar K^o(0) .
$$
\par
The probability that meson $K^o$ produced at moment $t = 0$ will
be at moment $t \neq 0$ in the state of $\bar K^o$ meson, is given
by a squared absolute value of the amplitude in (19), i.e.:
$$
P(K^o \rightarrow \bar K^o) = \mid(\bar K^o(0) \cdot K^o(t))\mid^2
= cos^2 \theta sin^2 \theta \cdot
$$
$$
[e^{-\Gamma_1 t} + e^{-\Gamma_2 t} - 2 e^{-\frac{(\Gamma_1 +
\Gamma_2) t}{2}} cos ((E_2 -E_1) t)] , \eqno(20)
$$
since $cos^2 \theta = sin^2 \theta = \frac{1}{2}$, then
$$
P(K^o \rightarrow \bar K^o) = \frac{1}{4}[e^{-\Gamma_1 t} +
e^{-\Gamma_2 t} - 2 e^{-\frac{(\Gamma_1 + \Gamma_2) t}{2}} cos
((E_2 -E_1) t)] , \eqno(21)
$$
and
$$
P(K^o \rightarrow  K^o) = P(\bar K^o \rightarrow  \bar K^o) =
$$
$$
= \frac{1}{4}[e^{-\Gamma_1 t} + e^{-\Gamma_2 t} + 2
e^{-\frac{(\Gamma_1 + \Gamma_2) t}{2}} cos ((E_2 -E_1) t)] ,
\eqno(22)
$$
where $E_2 - E_1$ is determined by expression (18).
\par
Figure 2 gives  transition probabilities (expr. (22)) of primary
$K^o$ meson  into $K^o$ ($K^o \to  K^o$) and Figure 3 shows
transition probabilities (expr. (21)) of primary $K^o$ meson into
$\bar K^o$ ($K^o \to \bar K^o$) in vacuum taking into account
$K^o_1, K^o_2$ decays in dependence on time. Presence of $K^o,
\bar K^o$ mesons at oscillations is detected via their decay into
$K^o \to \pi^{-} \nu_e e^{+}$ and $\bar K^o \to \pi^{+} e^{-} \bar
\nu_e$.

\begin{figure}[h!]
\begin{center}
\includegraphics[width=8.0cm]{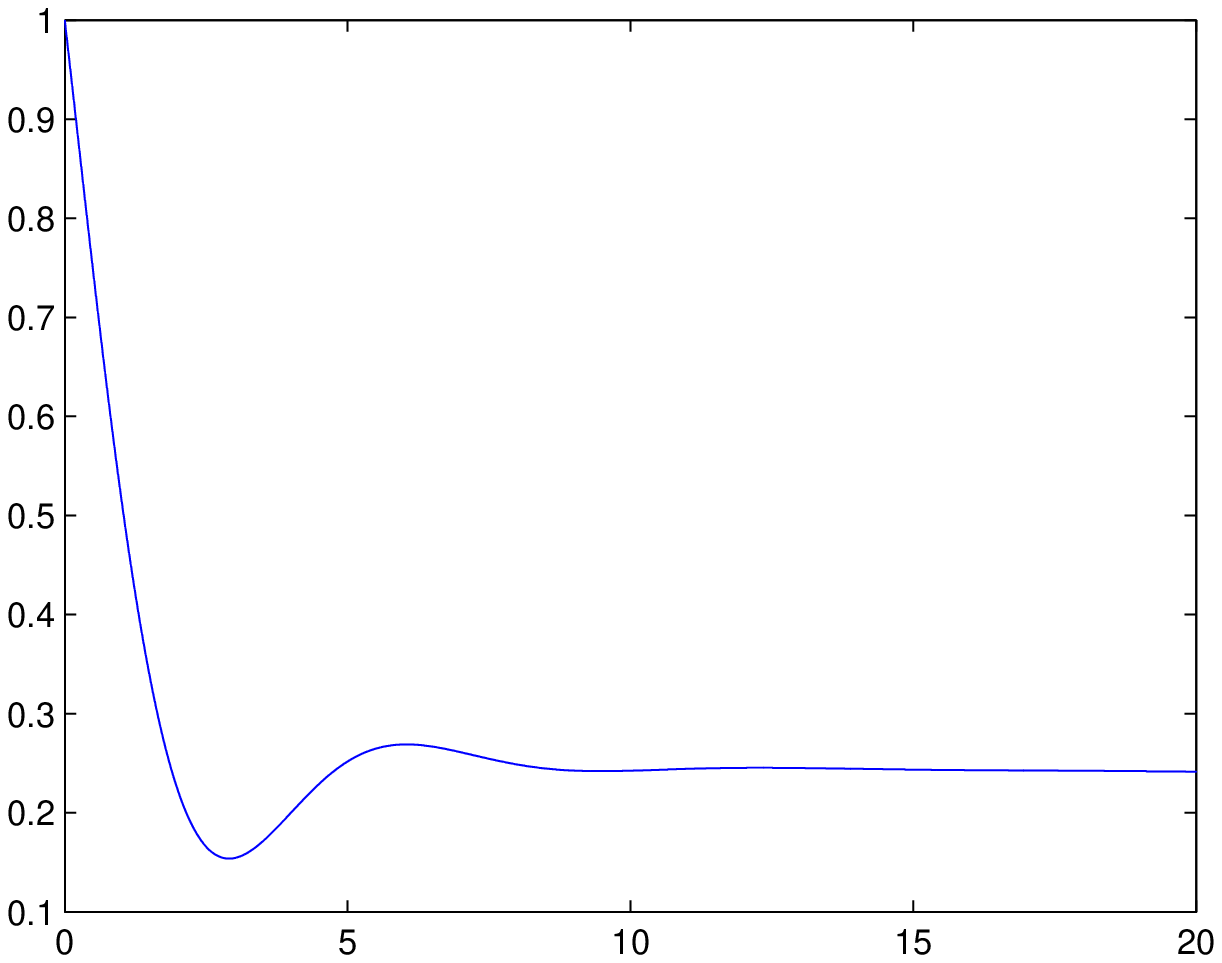}
   \end{center}
   \end{figure}
Figure 2. Transition probabilities (expr. (22)) of primary $K^o$
meson into $K^o$ ($K^o \to K^o$) in vacuum taking into account
$K^o_1, K^o_2$ decays in dependence on time.

\newpage
\begin{figure}[h!]
\begin{center}
\includegraphics[width=8.0cm]{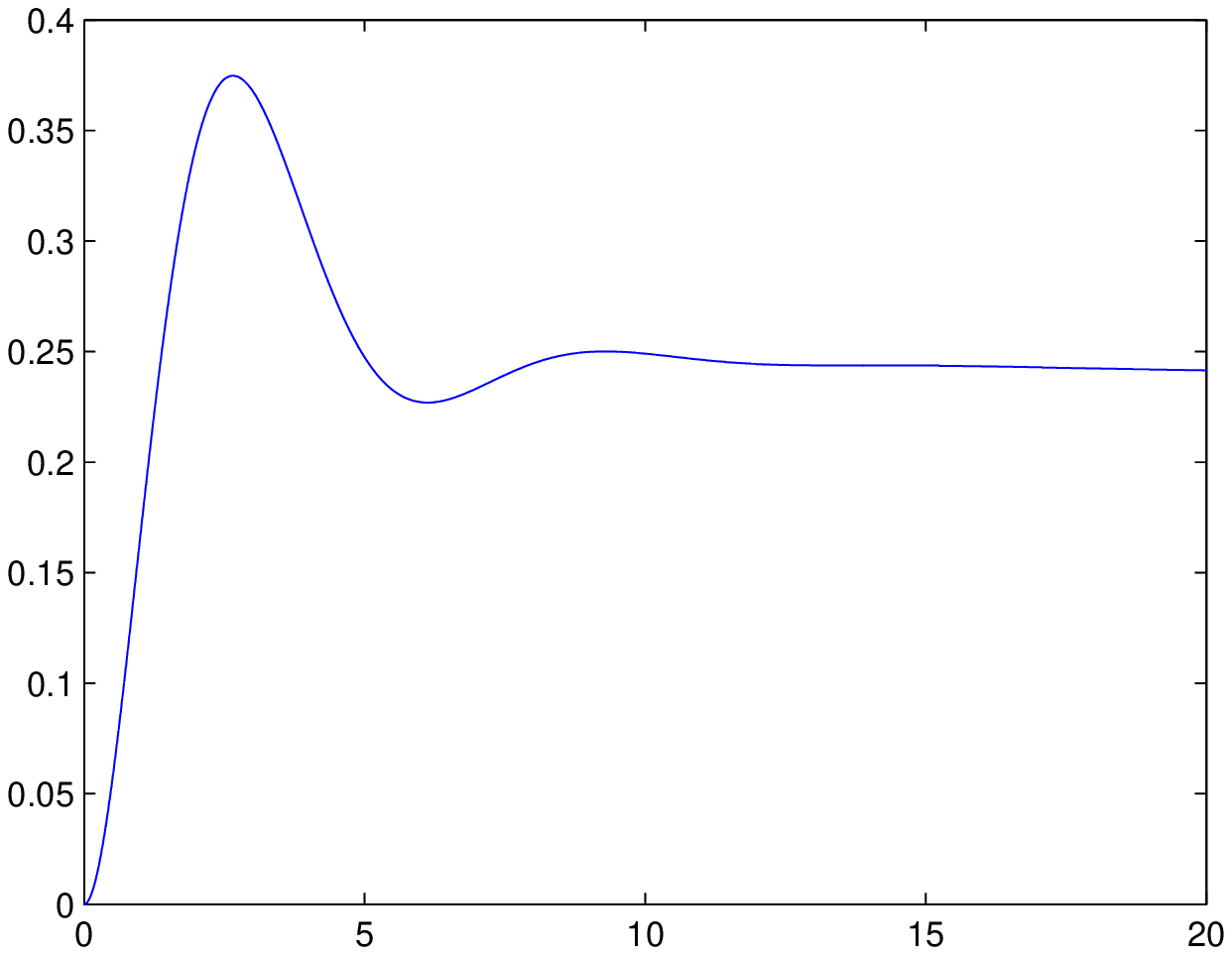}
   \end{center}
   \end{figure}
Figure 3. Transition probabilities (expr. (21)) of primary $K^o$
meson into $K^o \to \bar K^o$ in vacuum taking into
account $K^o_1, K^o_2$ decays in dependence on time.\\

\par
From expressions (21), (22) and Figures 2, 3 we see that at big
distances the long living terms $\frac{1}{4} e^{-\Gamma_2 t}$
remain. The sum of these two terms is $\frac{1}{2} e^{-\Gamma_2
t}$. It means that at long distances this term is present. It is
interesting to know how this process arises at oscillations. From
expressions (21), (22) we see that since $\Gamma_1
>> \Gamma_2$ then the term $e^{-\Gamma_1 t}$ will vanish much faster than the term
$e^{-\Gamma_2 t}$. Since oscillations between $K^o, \bar K^o$ can
be realized only between equal portions of numbers of
$e^{-\Gamma_1 t}$ and $e^{-\Gamma_2 t}$ then residuary without
partner $K^0_2$ mesons cannot participate in oscillations and will
slowly decay.
\par
The same result can be obtained using expressions(5) and (17),
then
$$
K^o(t) = \frac{1}{\sqrt{2}}( e^{-i E_1 t - {\frac{\Gamma_1}{2}} t}
K^o_1(0) +
 e^{-i E_2 t - {\frac{\Gamma_2}{2}} t} K^o_2(0)),
$$
$$
 \bar K^o(t) = \frac{1}{\sqrt{2}}( e^{-i E_1 t- {\frac{\Gamma_1}{2}} t} K^o_1(0) -
 e^{-i E_2 t- {\frac{\Gamma_2}{2} t}} K^o_2(0)) .  \eqno(23)
$$
From expressions in (5) we see that at big distances ($t \gg
\frac{1}{\Gamma_1}$) since $\Gamma_1 \gg \Gamma_2$ we obtain that
$$
K^o(t \gg \frac{1}{\Gamma_1}) \simeq \frac{1}{\sqrt{2}}(e^{-i E_2
t - {\frac{\Gamma_2}{2}} t} K^o_2(0)),
$$
$$
\bar K^o(t \gg \frac{1}{\Gamma_1}) \simeq \frac{1}{\sqrt{2}}(-
 e^{-i E_2 t- {\frac{\Gamma_2}{2} t}} K^o_2(0)) .  \eqno(24)
$$
In reality at big distances the $K^o_2(0)$ states remain and there
are no $K^o, \bar K^o$ although in the above transition
probabilities we have to see these mesons. Therefore we have to
come to a conclusion that at big distances the standard
expressions for transition probabilities obtained in the framework
of the oscillation theory are incorrect in the case when meson
decays are present. This is the main result of this work. The
author would like to draw the attention to that.
\par
For clearness, as an example, below we give two Figures (4 and 5)
for transition probabilities where the expression for a
probability of $K^o_2$ decay is included.
\par
Figure 4 gives transition probabilities  of primary $K^o$ meson
into $\bar K^o$ ($K^o \to \bar K^o$) together with the expression
for $K^o_2$ meson decay probability. And Figure 5 gives transition
probabilities of primary $K^o$ meson into $K^o$ ($K^o \to \bar
K^o$) in vacuum taking into account $K^o_1, K^o_2$ decays in
dependence on time and expression for $K^o_2$ meson decay
probability. It is implied that a real probability of $K^o \to
K^o, \bar K^o$ transitions is obtained after subtraction of the
long living tail.

\begin{figure}[h!]
\begin{center}
\includegraphics[width=8cm]{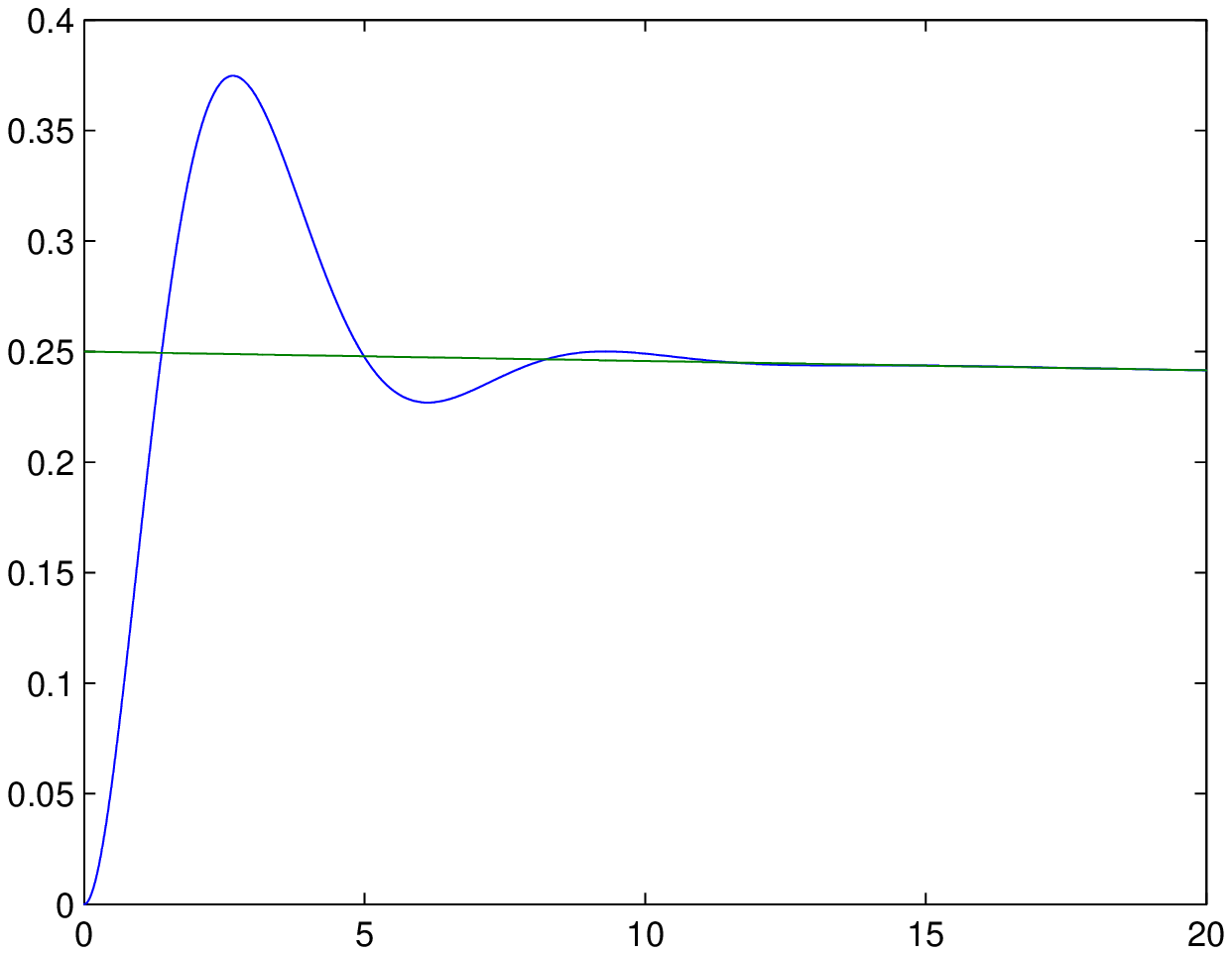}
   \end{center}
   \end{figure}
Figure 4. Transition probabilities of the primary $K^o$ meson into
$\bar K^o$ in vacuum taking into account $K^o_1, K^o_2$ decays and
expression for $K^o_2$ decay in dependence on time.

\begin{figure}[h!]
\begin{center}
\includegraphics[width=8cm]{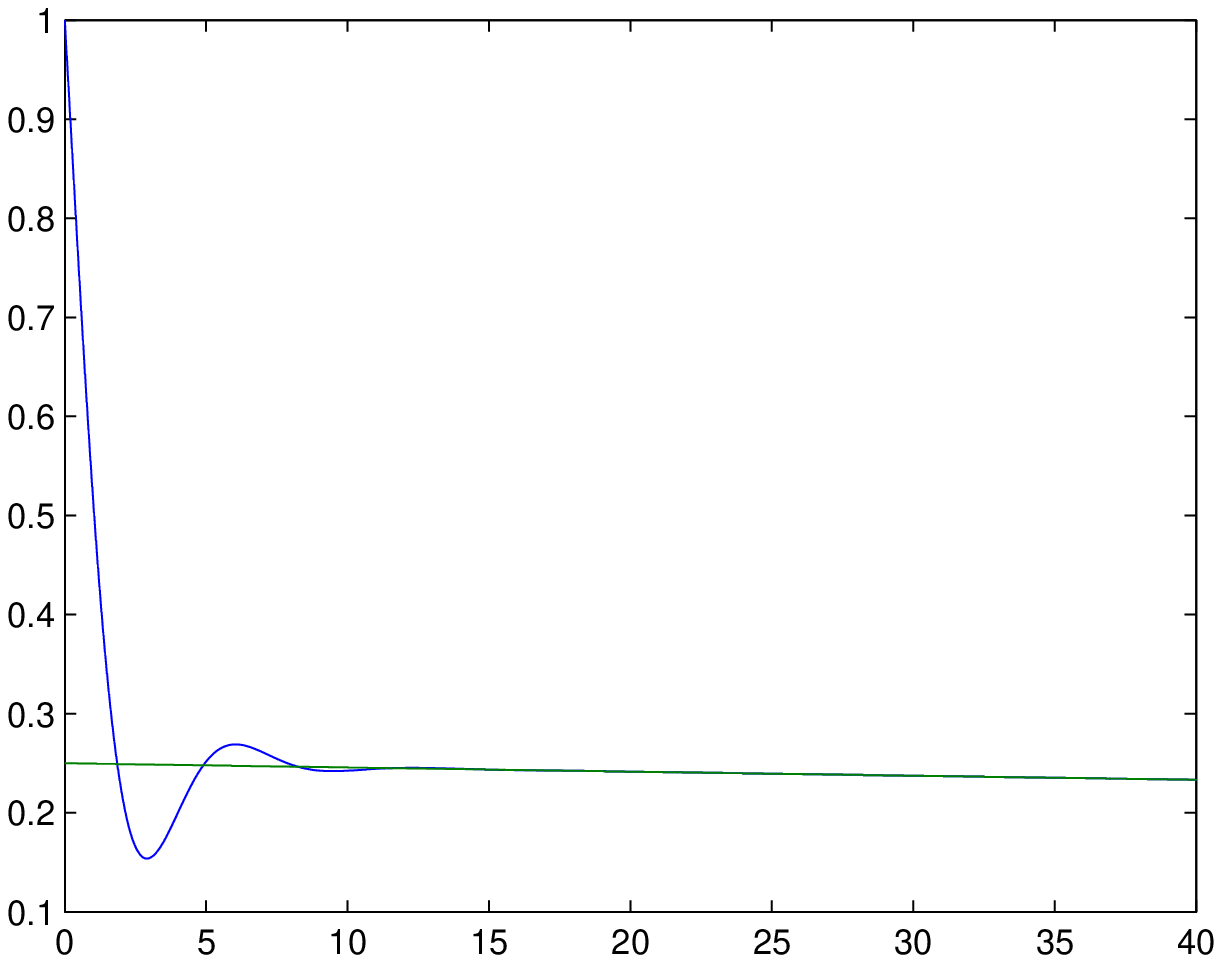}
   \end{center}
   \end{figure}
Figure 5. Transition probabilities of the primary $K^o$ meson into
$K^o \to \bar K^o$ in vacuum taking
into account $K^o_1, K^o_2$ decays in dependence on time.\\

\section{Conclusion}

This work was devoted to consideration of $K^o$ meson oscillations
at strangeness violation in weak interactions via $K^o_1, K^o_2$
mesons in two cases - without taking meson decays into account and
with them. It is noted that in the first case the theory of $K^o$
oscillations works well while describing $K^o \to K^o, \bar K^o$
and $\bar K^o \to K^o, \bar K^o$ meson transitions. But in the
second case, when we take $K^o_1, K^o_2$ meson decays into
account, the expression for transition probabilities $K^o \to K^o,
\bar K^o$ and $\bar K^o \to K^o, \bar K^o$ mesons does not work
correctly since at long distances indeed very small $K^o_1$ mesons
 remain and then their oscillations are absent. At long distances
 mainly $K^o_2$ mesons are present and the condition for oscillations
 to appear is absent.
\par
So, at long distances in this process long living $K^0_2$ mesons
but not $K^o, \bar K^o$ mesons in contrast to standard expressions
(20)-(22), are present. That was confirmed by the experiments
\cite{8, 9, 13}. The main result of this work is to stress this
feature of the theory of oscillations.  It is necessary to
emphasize that the presence of $K^o, \bar K^o$ mesons at
oscillations is detected via their decay into $K^o \to \pi^{-}
\nu_e e^{+}$ and $\bar K^o \to \pi^{+} e^{-} \bar \nu_e$. Also it
is necessary to stress that for presence of $CP$ violation in the
weak interactions $K^0_1, K^0_2$ mesons transformed into
superposition states of $K_S, K_L$ mesons.
\par
The analogues analysis can be also used while considering $CP$
violation in the weak interactions.

\end{document}